\documentclass[prd,showpacs,twocolumn,superscriptaddress,floatfix,nofootinbib,10pt]{revtex4-2}

\usepackage{setspace}
\usepackage[utf8]{inputenc}
\usepackage{float}
\usepackage{amsmath,amssymb,amsfonts,bm}
\usepackage{graphicx}
\usepackage{multirow}
\usepackage[usenames,dvipsnames]{color}
\allowdisplaybreaks
\usepackage[
colorlinks=true,
linkcolor=blue,
breaklinks=true,
urlcolor=blue,
citecolor=blue]{hyperref}

\usepackage{orcidlink}
\usepackage{subfigure}
\usepackage{soul}
\usepackage{xcolor}
\usepackage{physics}
\usepackage{booktabs}
\usepackage{ulem}

\definecolor{green}{rgb}{0,0.6,0}
\newcommand{\mev}{\textrm{ MeV}}

\newcommand{\GXNU}{\affiliation{Department of Physics, Guangxi Normal University, Guilin 541004, China}}

\newcommand{\GXZD}{\affiliation{Guangxi Key Laboratory of Nuclear Physics and Technology, Guangxi Normal University, Guilin 541004, China}}

\newcommand{\CSU}{\affiliation{School of Physics, Central South University, Changsha 410083, China}}

\newcommand{\IFIC}{\affiliation{Departamento de F\'{\i}sica Te\'orica and IFIC, Centro Mixto Universidad de
Valencia-CSIC Institutos de Investigaci\'on de Paterna, Apartado 22085,
46071 Valencia, Spain}}

\newcommand{\UCAS}{\affiliation{School of Physics, University of Chinese Academy of Sciences, Beijing 100049, China}}

\newcommand{\ZZU}{\affiliation{School of Physics, Zhengzhou University, Zhengzhou 450001, China}}

\newcommand{\SCNT}{\affiliation{Southern Center for Nuclear-Science Theory (SCNT), Institute of Modern Physics,
Chinese Academy of Sciences, Huizhou 516000, China}}

\begin{document}
\title{How to unravel the nature of the $\Sigma^*(1430) (1/2^-)$ state from correlation functions}

\begin{abstract}
	We calculate the correlation functions for the $\bar K^0  p, \pi^+ \Sigma^0, \pi^0 \Sigma^+, \pi^+ \Lambda$, and $\eta \Sigma^+$ states, which in the chiral unitary approach predict an excited $\Sigma^*(1/2^-)$ state at the $\bar K N$ threshold, recently observed by the Belle Collaboration. Once this is done, we tackle the inverse problem of seeing how much information one can obtain from these correlation functions. With the resampling method, one can determine the scattering parameters of all the channels with relative precision by means of  the analysis in a general framework, and find a clear cusplike structure corresponding to the $\Sigma^*(1/2^-)$ in the different amplitudes at the $\bar{K}N$ threshold.
\end{abstract}

\author{Hai-Peng Li}%
\GXNU%

\author{Chu-Wen Xiao}%
\email{xiaochw@gxnu.edu.cn}
\GXNU%
\GXZD%
\CSU%

\author{Wei-Hong Liang}%
\email[Corresponding author: ]{liangwh@gxnu.edu.cn}
\GXNU%
\GXZD%

\author{Jia-Jun Wu}%
\email{wujiajun@ucas.ac.cn}
\UCAS%
\SCNT%

\author{En Wang}%
\email{wangen@zzu.edu.cn}
\ZZU%
\GXZD%

\author{Eulogio Oset}%
\email{Oset@ific.uv.es}
\IFIC%
\GXNU%

\maketitle

\section{Introduction}\label{sec:Intr}
The $\Sigma^*$ states with $J^P=1/2^-$ are a benchmark for hadron spectroscopy, usually demanding for
their interpretation large meson-baryon components on top of the conventional $qqq$ seed, when not being dominated by this meson-baryon cloud.
A recent review on this topic can be found in Ref.~\cite{Wang:2024jyk}.
In the present work we are concerned just about one of these states that still is controversial, a $\Sigma^*(1/2^-)$ state appearing around the $\bar K N$ threshold, with mass $1430\; \rm MeV$ and relatively narrow width.
This state is predicted in studies of meson-baryon interaction using the chiral unitary approach, where the two $\Lambda(1405)$ states appear \cite{Kaiser:1995eg,Oset:1997it}, but it is quite elusive and it was missed in early investigations.
A clear state as a pole in the second Riemann sheet was found in the work of Ref.~\cite{Oller:2000fj}.
This was corroborated in Ref.~\cite{Jido:2003cb}, where using a slightly different input in the same chiral unitary approach one failed to get this $\Sigma^*(1/2^-)$ state when using physical masses of the states, but obtained it with masses slightly changed to be closer to the SU(3) limit of equal masses for the octet baryons and equal masses for the octet mesons.
The two pole structure remains when constraints on $\pi N$ and $KN$ data are considered together with the data of $\bar K N$, and the pole positions can be determined with better precision \cite{Lu:2022hwm}.
Actually in the same work a $\Sigma^*$ state appears at $1432 \mev$, as in the former quoted works.
We anticipate here that while there is a drastic step between having or not having the pole in a certain Riemann sheet, the physical magnitudes around the $\bar K N$ threshold change continuously and smoothly from one situation to the other.
Whether disclosed or not, and having the state as a bound or a virtual state, all different works along the line of the chiral unitary approach produce this state and have some threshold enhancement in the scattering amplitudes of the coupled channels around the $\bar K N$ threshold in isospin $I=1$ \cite{Oset:2001cn,Khemchandani:2018amu,Kamiya:2016jqc,Oller:2006jw,Garcia-Recio:2002yxy,Lutz:2001yb,Guo:2012vv}.
The $I=1$ state around the $\bar K N$ threshold remains when the mixing of pseudoscalar-baryon and vector-baryon components is allowed \cite{Khemchandani:2012ur,Khemchandani:2011mf}.
The data from $\pi \Sigma$ photoproduction in $\gamma p \to K\Sigma \pi$ \cite{CLAS:2013rjt} were considered in Ref.~\cite{Roca:2013cca} to constrain the parameters of the theory and no pole was found in the second Riemann sheet, although a cusplike structure appeared in $|T|^2$ for the different amplitudes at the $\bar K N$ threshold.

The situation of this state is similar to the one of the $a_0(980)$, which also appears as a clear cusp in recent experiments \cite{CLEO:2004umu,BESIII:2016tqo,Belle:2020fbd,BESIII:2020pxp,BESIII:2023htx} and theoretical studies \cite{Oller:1998hw,Xie:2014tma,Liang:2016hmr,Toledo:2020zxj,Ikeno:2024fjr}.
In spite of this clear situation, the $a_0(980)$ is accepted as a resonance, and, correspondingly, there should not be a problem in calling this $\Sigma^*(1/2^-)$ state a resonance close to the $\bar K N$ threshold, a criterium that we will adopt in what follows.

Many suggestions of experiments have been done to find this elusive state.
In Ref.~\cite{Lyu:2023oqn} the $\gamma n \to K^+ \Sigma^{*-}(1/2^-)$ reaction is proposed to observe this state.
In Refs.~\cite{Ren:2015bsa,Wu:2013kla}  the $\bar\nu_l p \to  l^+\Phi B$ reaction, with $\Phi$ and $B$ being mesons and baryons of the SU(3) octet, is also shown to be suited to the observation of this state.
In Ref.~\cite{Wang:2015qta}, the authors proposed to search for the $\Sigma^*(1/2^-)$ in the $\Sigma \pi$ mass distribution of the process $\chi_{c0}(1P) \to \bar \Sigma \Sigma \pi$.
A similar work was performed in Ref.~\cite{Liu:2017hdx} suggesting the $\chi_{c0} \to \bar\Lambda \Sigma \pi$ decay, considering the contributions from
the $\pi \Sigma$ and $\pi \bar \Lambda$ final state interactions within the chiral unitary approach, from where the $\Lambda(1405)$ and the $\Sigma^*(1/2^-)$ around $\bar K N$ threshold emerged.
Also, the spectra of the Belle experiment on the $\Lambda^+_c \to pK^0_S \eta$  reaction \cite{Belle:2022pwd}  was analyzed in Ref.~\cite{Li:2024rqb}, showing that a fit to the data improved substantially by including the contribution of a $\Sigma^*(1/2^-)$ state at $1380\, \rm MeV$.
In Ref.~\cite{Xie:2018gbi}, the  $\Lambda_c^+ \to \pi^+ \pi^0 \pi^- \Sigma^+$ reaction was shown to be driven by a triangle singularity that reinforced the production of the $\Sigma^*(1430)$ state.
In the same work, it was suggested to look at the $\Lambda_c^+ \to  \pi^+ \pi^+ \pi^- \Lambda$ reaction, where the signal should be equally seen.
This reaction has been performed recently by the Belle Collaboration \cite{Belle:2022ywa}  and a clear signal is seen at the $\bar K N$ threshold, both in the $\Lambda \pi^+$ and $\Lambda \pi^-$ mass distributions.
This is the first clear evidence of the existence of this state.
The authors of the work also state that from their analysis they cannot discriminate from the peak being due to a resonance or to a cusp in the $\bar K N$ threshold.
It should be clear from the former discussion that we also do not make a strong difference between the two scenarios since one can pass from one to the other with a minor change in the strength of the interaction or other parameters of the theory.
Yet, the nature of this state and its properties deserve further investigations. This is the purpose of the present work.

In this work, we propose to use the femtoscopic correlation functions of the coupled channels  which generate the $\Sigma^*(1430)$ state in the chiral unitary approach, to further reveal the origin and properties tied to the state.
Thus we construct the correlation functions for the channels $\bar K^0 p, \pi^+ \Sigma^0, \pi^0 \Sigma^+, \pi^+ \Lambda$ and $\eta \Sigma^+$, and in a second step we face the inverse problem in which, starting from these correlation functions we determine the possible existence of an $I=1$ state close to the $\bar K N$ threshold, and the scattering length and effective range for the different channels.
While the procedure might look like a tautology, this is not the case because the information on the correlation functions, which is related to amplitudes above the threshold of the channels, is more limited than the one of the original model from where they are evaluated.
Then it is important to know how much information one can get from the correlation functions and the uncertainties expected for the magnitudes determined from them assuming certain errors in the measurements of the correlation functions.

The nature of the hadronic states, as standard quark model states, molecular states of other hadronic components, multiquark states or others, is an issue of permanent debate \cite{Esposito:2014rxa,Guo:2017jvc}.
In the chiral unitary approach, the two $\Lambda(1405)$ states appear as molecular states, stemming from the interaction of the components of the coupled channels.
They do not correspond to three quark states, or compact multiquark states.
This is also the case of the $\Sigma^*(1430)$.
Should one of these states correspond to a three quark state with no connection to the $\pi \Sigma$ or $\bar K N$ channels, the correlation functions of these components would not shed light on the states.
On the contrary, if the states are molecular states, hence being formed mostly from these components, the correlation functions of these channels would contain information on their possible bound states.
In particular, the bound states close to a threshold of two hadrons have a large chance of being a molecule of these hadrons \cite{Dong:2020hxe}.
Yet, as discussed in Refs.~\cite{Dai:2023kwv,Song:2023pdq}, it is still possible to have these states corresponding to nonmolecular states, but at the price of having a very small scattering length and a huge effective range.
With the correlation functions one can also determine these scattering observables and come with strong conclusions concerning the nature of the states, as done in Ref.~\cite{Dai:2023kwv} concerning the $T_{cc}$ state and in Ref.~\cite{ Song:2023pdq} concerning the $X(3872)$ state.

In the present case one can argue as follows. 
In the extreme case that the $\Sigma(1430)$ state were purely nonmolecular, with no coupling to the meson-baryon components, the correlation functions, using the inverse method that we explain below, would not give any information on this state. 
Since we have experimental information on the existence of the state, we would already conclude that the state has no molecular components. 
Yet, even if small, a compact state will have some coupling to these meson-baryon components, in which case the inverse method that we use in this paper would trace the existence of the pole. 
What can we say in this case about its nature? 
One possibility is to calculate the compositeness of the state, via $-g_i^2\, \partial G_i/\partial E$, with $g_i$ the couplings and $G_i$ the meson-baryon loop functions \cite{Hyodo:2013nka,Hyodo:2011qc,Sekihara:2014kya,Gamermann2010}. 
Yet, with coupled channels this magnitude is generally complex and cannot be interpreted as a probability (see a particular case for the two $\Lambda(1405)$ resonances in Ref.~\cite{Sekihara:2014kya}). 
Yet, it still provides some ideas of the strength of a given component in the wave function \cite{Aceti:2014ala}, such that some information can be obtained from there, in particular this compositeness would be found very small compared to unity in the case of a clearly nonmolecular state. 
Another way would be to evaluate the scattering length and effective range from this correlation function, as we will explain in the paper, and see the results obtained. 
An abnormally large negative effective range for the threshold $\bar K N$ component, of tens of fm or larger would be an indication of a mostly composite state, as discussed in detail in Refs.~\cite{Dai:2023kwv,Song:2023pdq}. 
The combination of both pieces of information can tell us much about the nature of the state. 
And then, since so far the chiral unitary approach is providing a reasonable description of this state (it was predicted before its observation), any substantial deviation of the results obtained from the correlation functions from those obtained here with the chiral unitary approach, in particular for the scattering length and effective range of the $\bar K N$ channel, would also prove that the nature of that state is rather different from the picture provided by the chiral unitary approach. 

We also take advantage here to clarify what we mean by inverse problem.
This is done in contrast to the direct problem that we define as constructing the correlation functions from a theoretical model.
The inverse problem would consist in determining the parameters of the theoretical model from the correlation functions data.
Yet, we do not make a specific model but use the simplest and most general form that can account for the interaction of the coupled channels in a limited range of energies.
We will refer to this method as ``minimal model dependent".
We do not pay much attention to the parameters obtained for this model, being conscious that there are strong correlations among them, and directly evaluate observables and their uncertainties.
Hence, by using this minimal model dependent method, we go from the correlation functions to the determination of scattering lengths and effective ranges, plus the search of possible existing bound states or resonances.

Femtoscopic correlation functions are emerging as a relevant experimental tool, showing the potential to provide information on the interaction of hadrons, scattering parameters and the nature of hadronic resonances.
Experimental measurements are already available in Refs.~\cite{ALICE:2017jto,ALICE:2018ysd,ALICE:2019hdt,ALICE:2019eol,ALICE:2019buq,ALICE:2019gcn,ALICE:2020mfd,ALICE:2021szj,ALICE:2021cpv,ALICE:2022enj,STAR:2014dcy,STAR:2018uho,Fabbietti:2020bfg,Feijoo:2024bvn} and theoretical studies are also progressing on the same path \cite{Morita:2014kza,Ohnishi:2016elb,Sarti:2023wlg,Morita:2016auo,Hatsuda:2017uxk,Mihaylov:2018rva,Haidenbauer:2018jvl,Morita:2019rph,Kamiya:2019uiw,Kamiya:2021hdb,Kamiya:2022thy,Liu:2023uly,Vidana:2023olz,Albaladejo:2023pzq,Liu:2023wfo,Liu:2022nec,Liu:2023wfo,Torres-Rincon:2023qll,Li:2023pjx,Molina:2023oeu,Li:2024tof,Liu:2024uxn}, with some later works involved in a model independent analysis of experimental data to extract the encoded information on scattering parameters and possible related bound states and resonances \cite{Ikeno:2023ojl,Albaladejo:2023wmv,Feijoo:2023sfe,Molina:2023jov} .
In the present work, we explore the potential of using correlation functions to study the $\Sigma^*(1430)$, find out about its nature, and at the same time determine scattering parameters for the related coupled channels $\bar K^0 p, \pi^+ \Sigma^0, \pi^0 \Sigma^+, \pi^+ \Lambda$, and $\eta \Sigma^+$.

\section{Formalism}
\label{sec:Form}
\subsection{Summary of the chiral unitary approach for $\bar KN$ in coupled channels}
\label{subsec:chua}
In this work, we first provide a brief introduction of the $\bar{K}N$ and related channel interactions within the chiral unitary approach~\cite{Oset:1997it}. 
The work relies on the use of the lowest order chiral Lagrangian with unitarity in the coupled channels.
While much progress using next to leading order Lagrangians (see Ref.~\cite{Lu:2022hwm} and references within) has been done later, it is well known that the lowest order approach, with just one free parameter, provides a very good reproduction of the experimental data \cite{Oset:1997it}.

To avoid using the Coulomb interaction, we will take the coupled channels with $I_{3}=1$ which only have one charged particle.
The channels are $\bar{K}^{0}p$, $\pi^{+}\Sigma^{0}$, $\pi^{0}\Sigma^{+}$, $\pi^{+}\Lambda$, and $\eta\Sigma^{+}$ that we label $1, 2, 3, 4, 5$ respectively.
For the purpose of getting the $\Sigma^{*}(1430)$ state, 
we can safely ignore the $K\Xi$ channel, the threshold of which is far away from $1430 \mev$.
Considering the phase convention of the isospin multiplets: $(\bar{K}^{0}, -K^{-})$, $(p, n)$, $(-\pi^{+}, \pi^{0}, \pi^{-})$ and $(-\Sigma^{+}, \Sigma^{0},\Sigma^{-})$, 
\footnote{Our phase convention is such that the $q_i \bar q_j$ matrix of quarks expressed in terms of pseudoscalar mesons of the octet corresponds to the usual SU(3) matrix $\mathcal{P}$ used in chiral perturbation theory \cite{Oset:1997it}, thus $K^-=s\bar u, \bar K^0 =s\bar d$, and the same prescription is used for the octet of baryons.}
we can easily convert isospin states to charge states, and the relationship is given by
\begin{align}
	\label{eq:c2i}
	\ket{\pi^{+}\Sigma^{0}}=&-\frac{1}{\sqrt{2}}\big(\ket{\pi\Sigma,I=2,I_{3}=1}+\ket{\pi\Sigma,I=1,I_{3}=1}\big),\nonumber\\
	\ket{\pi^{0}\Sigma^{+}}=&-\frac{1}{\sqrt{2}}\big(\ket{\pi\Sigma,I=2,I_{3}=1}-\ket{\pi\Sigma,I=1,I_{3}=1}\big),\nonumber\\
	\ket{\bar{K}^{0}p}=&\ket{\bar{K}N,I=1,I_{3}=1},\nonumber\\
	\ket{\pi^{+}\Lambda}=&-\ket{\pi\Lambda,I=1,I_{3}=1},\nonumber\\
	\ket{\eta\Sigma^{+}}=&-\ket{\eta\Sigma,I=1,I_{3}=1}.
\end{align}

The interaction of the coupled channels is given by
\begin{equation}
	\label{eq:vij}
	V_{ij}=-\frac{1}{4f^2}\, C_{ij}\,(k_{i}^{0}+k_{j}^{0}),\quad f=93 \mev,
\end{equation}
with
\begin{equation}
k_{i}^{0}=\frac{s+m_{i}^{2}-M_{i}^{2}}{2\sqrt{s}},
\end{equation}
where $\sqrt{s}$ is the c.m. energy of the meson-baryon system, and $m_{i}$ and $M_{i}$ are the masses of the meson and baryon in channel $i$, respectively.
The coefficients $C_{ij}$ of Eq.~\eqref{eq:vij} are given in Table \ref{tab:cij}.
\begin{table*}[htbp]
\centering
 \caption{The values of $C_{ij}$ coefficients of  different channels in the charge basis.}
 \label{tab:cij}
\setlength{\tabcolsep}{22pt}
\begin{tabular}{c|ccccc}
\hline
\hline
$C_{ij}$ &   $\bar{K}^{0}p$ &  $\pi^{+}\Sigma^{0}$ & $\pi^{0}\Sigma^{+}$& $\pi^{+}\Lambda $ & $ \eta \Sigma^{+}$ \\
\hline
$\bar{K}^{0}p$&1 & $\frac{1}{\sqrt{2}}$ & $-\frac{1}{\sqrt{2}}$ &$ \sqrt{\frac{3}{2}}$ & $\sqrt{\frac{3}{2}}$ \\
$\pi^{+}\Sigma^{0}$& &$ 0 $&$ -2 $&$ 0 $&$ 0 $\\
$\pi^{0}\Sigma^{+}$& &  &$ 0 $&$ 0 $&$0$ \\
$\pi^{+}\Lambda	$& &  &  & $0$ & $0$ \\
$\eta\Sigma^{+}$& &  &  &  & $0$ \\
\hline
\hline
   \end{tabular}
\end{table*}
The scattering matrix $T$ follows the Bethe-Salpeter equation in coupled channels,
\begin{equation}
	\label{eq:tij}
	T=[I-VG]^{-1}V,
\end{equation}
where $G$ is a diagonal matrix with its elements $G_{i}$ being the meson-baryon loop function regularized with a cutoff. Following Ref.~\cite{Oller:1998hw}, $G_{i}$ is given by
\begin{align}
	\label{eq:gi}
	G_{i}=2M_{i}\int_{\abs{\vec{q}}<q_\text{max}}\frac{\dd^3q}{(2\pi)^3}\frac{w_{1}(\vec{q}\,)+w_{2}(\vec{q}\,)}{2\,w_{1}(\vec{q}\,)\,w_{2}(\vec{q}\,)}\nonumber\\
	\times\frac{1}{s-[w_{1}(\vec{q}\,)+w_{2}(\vec{q}\,)]^2+i\epsilon},
\end{align}
with $w_{1}(\vec{q}\,)=\sqrt{\vec{q}\,^2+m_{i}^2}$, $w_{2}(\vec{q}\,)=\sqrt{\vec{q}\,^{2}+M_{i}^2}$ and $q_{\text{max}}=630 \mev$ \cite{Oset:1997it}.

With this input, we obtain a pole at
\begin{equation}\label{eq:chuasp}
	\sqrt{s}_{p}=1431.83-i104.75\ \text{MeV}.
\end{equation}
The width is very large, but we shall discuss later the meaning of this result.

\subsection{Correlation functions}

The correlation functions of pairs of particles are obtained experimentally from the ratio of the probability of finding this pair with a relative momentum $p$, and the uncorrelated probability of finding the pair obtained usually with the mixed event method \cite{Lisa:2005dd}.
Under certain approximations it can be written as \cite{Morita:2014kza}
\begin{equation}\label{eq:Cp}
	C(p)= \int \dd^3 r \, S_{12}(\vec r\,) \, |\psi(\vec r, p)|^2,
\end{equation}	
where $\psi(\vec r, p)$	is the wave function of the pair of particles and $S_{12}(\vec r\,)$ is the source function
\begin{equation}
	S_{12}(\vec r\,)=\frac{1}{(\sqrt{4\pi}R)^3}\exp(-\frac{\vec r^{\,2}}{4R^2}).
\end{equation}
The source function $S_{12}(\vec r\,)$ defines the probability density of finding the two hadrons of the emitted pair at a given relative distance $\vec r$.
The parametrization of $S_{12}(\vec r\,)$ as a Gaussian, while containing approximations to a more complex situation, has proved empirically suited to study the correlation functions \cite{Morita:2014kza}.
The correlation functions are measured from studies of $pp, p{\rm A, AA}$ collisions at very high energies, of the order of $10 \, {\rm TeV}$. 
It is clear that in these collisions there are many other particles created at the same time of the creation of the investigated pair.
In Ref.~\cite{Albaladejo:2024lam}, a discussion is conducted on the relationship of the measurements leading to the correlation functions and the production experiments looking at mass distributions.
A link is established there between the source function and the form factor appearing for off shell propagation of particles.
Similarly, the idea that the source could be related to the distribution of total (not relative) momenta of the pairs, which determines the invariant mass of the rest of the particles obtained, is also hinted in Ref.~\cite{Albaladejo:2024lam}.
We anticipate progress in the future in understanding the precise meaning of the source function.
For the moment, whatever its origin, the phenomenological and experimental works quoted above give support to the use of the source function of the Gaussian type used here, with the parameter $R$, which as we shall see below, can be rather accurately determined from the data.

Following Ref.~\cite{Vidana:2023olz}, the correlation functions can be written as
\begin{align}\label{eq:C1}
	\mathcal{C}_{\bar{K}^{0}p}(p_{\bar{K}^{0}})=1&+4\pi\theta(q_{\text{max}}-p_{\bar{K}^{0}})\int\dd r \,r^2 \,S_{12}(r)\nonumber\\
						      &\times\Big\{\abs{j_{0}(p_{\bar{K}^{0}}r)+T_{11}(E)\,\tilde{G}_{1}(r,E)}^2\nonumber\\
						      &+\abs{T_{21}(E)\,\tilde{G}_{2}(r,E)}^2+\abs{T_{31}(E)\,\tilde{G}_{3}(r,E)}^2\nonumber\\
						      &+\abs{T_{41}(E)\,\tilde{G}_{4}(r,E)}^2+\abs{T_{51}(E)\,\tilde{G}_{5}(r,E)}^2\nonumber\\
						      &-j_{0}^2(p_{\bar{K}^{0}}r)\Big\},
\end{align}
\begin{align}\label{eq:CF2}
	\mathcal{C}_{\pi^{+}\Sigma^{0}}(p_{\pi^{+}})=1&+4\pi\,\theta(q_{\text{max}}-p_{\pi^{+}})\int\dd r \,r^2 \, S_{12}(r)\nonumber\\
						       &\times\Big\{\abs{j_{0}(p_{\pi^{+}}r)+T_{22}(E)\, \tilde{G}_{2}(r,E)}^2\nonumber\\
						      &+\abs{T_{12}(E)\,\tilde{G}_{1}(r,E)}^2+\abs{T_{32}(E)\,\tilde{G}_{3}(r,E)}^2\nonumber\\
						      &+\abs{T_{42}(E)\,\tilde{G}_{4}(r,E)}^2+\abs{T_{52}(E)\,\tilde{G}_{5}(r,E)}^2\nonumber\\
						      &-j_{0}^2(p_{\pi^{+}}r)\Big\},
\end{align}
\begin{align}\label{eq:CF3}
	\mathcal{C}_{\pi^{0}\Sigma^{+}}(p_{\pi^{0}})=1&+4\pi\theta(q_{\text{max}}-p_{\pi^{0}})\int\dd r \,r^2 \, S_{12}(r)\nonumber\\
						      &\times\Big\{\abs{j_{0}(p_{\pi^{0}}r)+T_{33}(E) \,\tilde{G}_{3}(r,E)}^2\nonumber\\
						      &+\abs{T_{13}(E)\,\tilde{G}_{1}(r,E)}^2+\abs{T_{23}(E)\,\tilde{G}_{2}(r,E)}^2\nonumber\\
						      &+\abs{T_{43}(E)\,\tilde{G}_{4}(r,E)}^2+\abs{T_{53}(E)\,\tilde{G}_{5}(r,E)}^2\nonumber\\
						      &-j_{0}^2(p_{\pi^{0}}r)\Big\},
\end{align}
\begin{align}\label{eq:CF4}
	\mathcal{C}_{\pi^{+}\Lambda}(p_{\pi^{+}})=1&+4\pi\, \theta(q_{\text{max}}-p_{\pi^{+}})\int\dd r \,r^2 \, S_{12}(r)\nonumber\\
						    &\times\Big\{\abs{j_{0}(p_{\pi^{+}}r)+T_{44}(E)\tilde{G}_{4}(r,E)}^2\nonumber\\
						      &+\abs{T_{14}(E)\,\tilde{G}_{1}(r,E)}^2+\abs{T_{24}(E)\,\tilde{G}_{2}(r,E)}^2\nonumber\\
						      &+\abs{T_{34}(E)\,\tilde{G}_{3}(r,E)}^2+\abs{T_{54}(E)\,\tilde{G}_{5}(r,E)}^2\nonumber\\
						      &-j_{0}^2(p_{\pi^{+}}r)\Big\},
\end{align}
\begin{align}\label{eq:CF5}
	\mathcal{C}_{\eta\Sigma^{+}}(p_{\eta})=1&+4\pi\theta(q_{\text{max}}-p_{\eta})\int\dd r \,r^2 \,S_{12}(r)\nonumber\\
						&\times\Big\{\abs{j_{0}(p_{\eta} r)+T_{55}(E)\,\tilde{G}_{5}(r,E)}^2\nonumber\\
						      &+\abs{T_{15}(E)\,\tilde{G}_{1}(r,E)}^2+\abs{T_{25}(E)\,\tilde{G}_{2}(r,E)}^2\nonumber\\
						      &+\abs{T_{35}(E)\,\tilde{G}_{3}(r,E)}^2+\abs{T_{45}(E)\,\tilde{G}_{4}(r,E)}^2\nonumber\\
						      &-j_{0}^2(p_{\eta}r)\Big\},
\end{align}
where $p_{i}$ are the momenta of the particles in the center of mass frame,
\begin{equation}\label{eq:pi}
	p_{i}=\frac{\lambda^{1/2}(s,m_i^2,M_i^2)}{2\sqrt{s}},
\end{equation}
and the $\tilde{G}_{i}(r,E)$ function is defined as
\begin{align}
	\tilde{G}_{i}=2M_{i}\int&\frac{\dd^3q}{(2\pi)^3}\frac{w_{1}(\vec{q}\,)+w_{2}(\vec{q}\,)}{2\,w_{1}(\vec{q}\,)\;w_{2}(\vec{q}\,)}\nonumber\\
	&\times\frac{j_{0}(\abs{\vec{q\,}}r)}{s-[w_{1}(\vec{q}\,)+w_{2}(\vec{q}\,)]^2+i\epsilon},
\end{align}
with $j_{0}$ the zeroth order spherical Bessel function. 
We use relativistic dynamics in the $T_{ij}$ and $\tilde{G}$ matrices.
The link to nonrelativistic approaches can be seen in detail in Ref.~\cite{Albaladejo:2024lam}.
In Eqs.~\eqref{eq:C1}-\eqref{eq:CF5}, the wave function of Eq.~\eqref{eq:Cp} has been written in terms of the $T$ and $\tilde{G}$ functions and automatically includes relativistic effects (see details in Ref.~\cite{Vidana:2023olz}).

As shown in detail in Ref.~\cite{Vidana:2023olz}, when one has a single channel, for instance channel $1$, in Eq.~\eqref{eq:C1} we would get the term $\{|j_{0}(p_{\bar{K}^{0}}r)+T_{11}(E)\tilde{G}_{1}(r,E)|^2 -j_{0}^2(p_{\bar{K}^{0}}r)$\}.
The contribution of the inelastic channels $|T_{j1}(E)\,\tilde{G}_{j}(r,E)|^2$ have, in general, a weight $W_j$ relative to $1$ of the observed channel, and one expects $W_j$ to be not much different from $1$ for other channels.
These weights can be determined from experiments as shown, for instance, in Ref.~\cite{Feijoo:2024bvn}.
In the present work, in order to see the power of the correlation functions to determine values of the observables and their uncertainties, we assume these weights as unity in the direct calculation and the inverse analysis.

\subsection{Observables}
We evaluate the scattering length $a$ and effective range $r_{0}$ for each channel using the relationship of the $T$ matrix of Eq.~\eqref{eq:tij} to that in the quantum mechanics formalism as described in Ref.~\cite{Gamermann2010}.
This relationship is given by
\begin{equation}
	T=-\frac{8\pi\sqrt{s}}{2M}f^{QM}\simeq-\frac{8\pi\sqrt{s}}{2M}\;\frac{1}{-\frac{1}{a}+\frac{1}{2}r_{0} \, p^2-ip},
\end{equation}
with $p$ given in Eq.~\eqref{eq:pi},
from which we easily find
\begin{equation}
	\frac{1}{a_{i}}=\left.\frac{8\pi\sqrt{s}}{2M_{i}} \; (T_{ii})^{-1}\right|_{\sqrt{s}_{\text{th},i}},
\end{equation}
\begin{equation}
	r_{i}=\frac{1}{\mu_{i}}\,\frac{\partial}{\partial \sqrt{s}}\left[\frac{-8\pi\sqrt{s}}{2M_{i}} \; (T_{ii})^{-1}+ip_{i}\right]_{\sqrt{s}_{\text{th},i}},
\end{equation}
where $\mu_{i}$ is the reduced mass of channel $i$, and $\sqrt{s}_{\text{th},i}$ is the threshold energy for the channel $i$.

\subsection{Inverse problem}
First, we use the bootstrap or resampling method \cite{Press1992,Efron1986,Albaladejo2016}, generating random centroids of the data with Gaussian distributed weights within the error bars of the correlation functions of the chiral unitary approach to which we associate $\pm 0.02$ error, typical of present correlation data.
We take $26$ points from each correlation function and assume these new centroids with the same errors of $0.02$.
We repeat the procedure about $50$ times and attempt to extract the maximum information available by using a general framework with minimal model dependence.
For the inverse problem, we assume that the general interaction between coupled channels has an energy dependence like Eq.~\eqref{eq:vij}, given by
\begin{equation}\label{eq:Vij}
	V_{ij}=-\frac{1}{4f^2}\, \tilde{C}_{ij}\,(k_{i}^{0}+k_{j}^{0}),
\end{equation}
and that this potential is isospin symmetric (isospin is slightly broken in the $T$ matrix due to the difference of masses between the particles in the same isospin multiplets and their effect on the loops).
These are the minimal requirements that we must demand to this coupled channels interaction.
Actually, since we are concerned only about the peak around $1430 \mev$, the term $k_{i}^{0}+k_{j}^{0}$ is practically constant and we can consider the approach nearly model independent.
Yet, a fair statement is to call the framework a ``minimal model dependent'' approach.
Using Eq.~\eqref{eq:c2i} we can have the new symmetric coefficients $\tilde{C}_{ij}$ shown in Table \ref{tab:cijnew}, where $\tilde{C}_{22}$ and $\tilde{C}'_{22}$ are coefficients of $I=1$ and $I=2$, respectively.
The fractional coefficients in the table indicate the $I=1$ component of the states according to Eq.~\eqref{eq:c2i}.
The $T$ matrix is given by Eq.~\eqref{eq:tij}, and the $G$ function from Eq.~\eqref{eq:gi} is used with a free parameter $q_{\text{max}}$.
We would have $11$ free parameters in $\tilde{C}_{ij}$ plus $q_\text{max}$ and $R$ to fit our five correlation functions.
\begin{table*}[htbp]
\centering
\caption{The value of $\tilde{C}_{ij}$ coefficients of  different channels in the charge basis.}
\label{tab:cijnew}
\setlength{\tabcolsep}{18pt}
\begin{tabular}{c|ccccc}
\hline
\hline
$\tilde{C}_{ij}$ &   $\bar{K}^{0}p$ &  $\pi^{+}\Sigma^{0}$ & $\pi^{0}\Sigma^{+}$& $\pi^{+}\Lambda $ & $ \eta \Sigma^{+}$ \\[2.5mm]
\hline
$\bar{K}^{0}p$&$\tilde{C}_{11} $&$ -\frac{1}{\sqrt{2}}\tilde{C}_{12} $&$ \frac{1}{\sqrt{2}}\tilde{C}_{12} $&$ -\tilde{C}_{14} $&$ -\tilde{C}_{15}$ \\[2.5mm]
$\pi^{+}\Sigma^{0}$&~&$ \frac{1}{2}(\tilde{C}_{22}+\tilde{C}'_{22}) $&$ \frac{1}{2}(-\tilde{C}_{22}+\tilde{C}'_{22}) $&$ \frac{1}{\sqrt{2}}\tilde{C}_{24} $&$ \frac{1}{\sqrt{2}}\tilde{C}_{25} $\\[2.5mm]
$\pi^{0}\Sigma^{+}$&~ & ~ &$ \frac{1}{2}(\tilde{C}_{22}+\tilde{C}'_{22}) $&$ -\frac{1}{\sqrt{2}}\tilde{C}_{24} $&$ -\frac{1}{\sqrt{2}}\tilde{C}_{25}$ \\[2.5mm]
$\pi^{+}\Lambda	$&~ & ~ & ~ &$\tilde{C}_{44}$ & $\tilde{C}_{45}$  \\[2.5mm]
$\eta\Sigma^{+}$&~ & ~ & ~ & ~ &$\tilde{C}_{55}$\\[2mm]
\hline
\hline
\end{tabular}
\end{table*}

The resampling method is particularly useful when one has many parameters and expects correlations between them.
In each fit of the resampling, a set of parameters is obtained and then the observables are evaluated.
After many such fits, the average of the observables and their dispersion are evaluated.
The values of the parameters can change from one fit to the other since different configurations of parameters can give rise to equivalent good fits to the data, but the relevant results are the observables and their uncertainties, which one obtains from the dispersion of the results with respect to their averages.

\section{Considerations on the $\Sigma^{*}(1430)$ state}
In the process of the resampling, we also search for possible poles of Eq.~\eqref{eq:tij} in the second Riemann sheet which is obtained using $G^{II}_i$ instead of $G_i$ of Eq.~\eqref{eq:gi} as
\begin{equation}
	\label{eq:g2}
	G_i^{II}(\sqrt{s})=G_i(\sqrt{s})+i\frac{M_i}{2\pi\sqrt{s}}\;q_{\text{on}},
\end{equation}
for $\Re(\sqrt{s})>M_i+m_i$, $q_{\text{on}}=\lambda^{1/2}(s,m_i^2,M_i^2)/2\sqrt{s}$, $\Im(q_{\text{on}})>0$. 
We refrain from making the change of Eq.~\eqref{eq:g2} to closed channels since $q_{\text{on}}$ becomes purely imaginary and Eq.~\eqref{eq:g2} increases the size of the negative $G_i(\sqrt{s})$ function, effectively increasing the size of attractive interactions and eventually producing states.
This procedure generates poles (virtual states). By adhering to Eq.~\eqref{eq:g2} without altering the potential, one approaches the results obtained from solving the equivalent Schr\"{o}dinger equation with a specified potential. In this equivalent Schr\"{o}dinger equation, there exists a threshold strength at which a bound state will vanish if the potential is slightly reduced.
An alternative approach is to consider Eq.~\eqref{eq:g2} also for closed channels.
In this case, when the procedure of Eq.~\eqref{eq:g2} does not produce a pole, the alternative method produces a pole in the second Riemann sheet of that channel (virtual pole).
In this procedure the pole moves smoothly from one Riemann sheet to the other.

With this former caveat when we perform the resampling procedure using Eq.~\eqref{eq:g2} we obtain poles, and we make the average and dispersion of the pole position and find
\begin{equation}
\label{eq:fitsp}
	\sqrt{s}_{p}=(1420\pm10)-i(101\pm19)\ \text{MeV}.
\end{equation}
This is highly surprising since it would imply a width of about $200 \mev$, when the width of the $\Sigma^{*}(1430)$ state is about $10-30 \mev$ according to the Belle experiment \cite{Belle:2022ywa}.
In Ref.~\cite{Roca:2013cca} one does not find the pole and instead one finds a cusp at the $\bar{K}N$ threshold for $\abs{T_{ij}}^2$ of the different amplitudes.
The result of Eq.~\eqref{eq:fitsp} is basically consistent with that of Eq.~\eqref{eq:chuasp}, and we learn that we can obtain the pole position from the correlation functions, with an uncertainty of about $10 \mev$  in the position and around $40 \mev$ in the width, with the assumed
uncertainty in the correlation functions.

\begin{figure}[t]
\begin{center}
\includegraphics[scale=0.55]{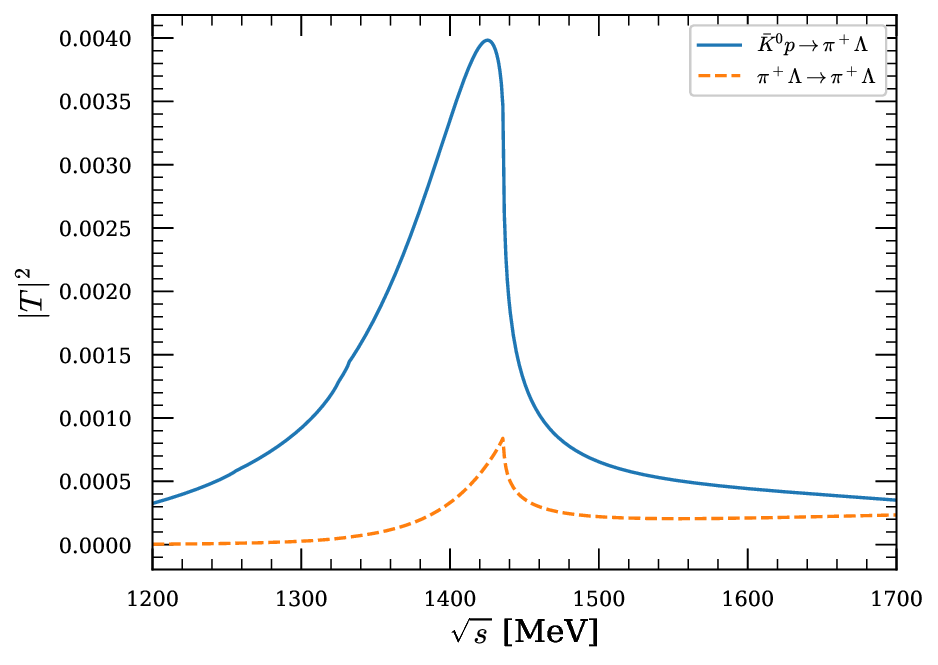}
\end{center}
\vspace{-0.7cm}
\caption{Absolute squared values of $T$ in $\bar{K}^{0}p\to\pi^{+}\Lambda$ and $\pi^{+}\Lambda\to\pi^{+}\Lambda$ of the chiral unitary approach taking $q_{\text{max}}=630$ MeV.}
\label{fig:t}
\end{figure}

The results of Eq.~\eqref{eq:chuasp} are similar to those obtained in Ref.~\cite{Oller:2000fj}, where, as shown in Ref.~\cite{Jido:2003cb}, a pole would appear close to the $\bar{K}N$ threshold and with a width of the order of $200 \mev$.
However, as discussed in Ref.~\cite{Jido:2003cb}, poles found around the threshold must be taken with caution and 
one should note that the experiment will see amplitudes in the physical real energy axis and in that case the extrapolation to the complex plane will be very model dependent.
This is actually the case here.
Indeed, in Fig.~\ref{fig:t} we show results for $\abs{T_{ij}}^2$ for different amplitudes, and we see that in all cases we find $\abs{T_{ij}}^2$ looking like a cusp around the $\bar K N$ threshold with an apparent width of $30-50 \mev$.

It is surprising that in spite of having a pole indicating a width of about $200 \mev$, the results of $\abs{T_{ij}}^2$ just produce a peak around the $\bar K N$ threshold with an apparent width of $30-50 \mev$, as seen in the Belle experiment.
This should serve as a caution when trying to interpret poles found around threshold.

It is also interesting to mention that in the work of Ref.~\cite{Roca:2013cca} the chiral potential was slightly modified and adjusted to the photoproduction data of Ref.~\cite{CLAS:2013rjt}.
This resulted in a reduction of the strength of the potentials by about a factor of the order of $0.8$.
In that case no pole was obtained for $I=1$, but a clear cusp at threshold resulted from the different amplitudes.
With this in mind, we make the exercise of reducing gradually the potential to see if the pole disappears.
We find that multiplying the $C_{ij}$ coefficients of Table \ref{tab:cij} by $0.97$, the pole already disappears according to Eq.~\eqref{eq:g2}.
This represents a discontinuous transition from possessing to lacking a pole. However, it is noteworthy that the amplitudes change continuously. In Fig.~\ref{fig:t}, the alterations caused by the normalization factor of $0.97$ are approximately $5\%$, with no apparent modifications in the shapes and widths of the structures depicted.
Actually, in the alternative method of using Eq.~\eqref{eq:g2} also for closed channels, the pole would smoothly move from one Riemann sheet to another, becoming a virtual pole.

\section{Results}
\begin{figure}[t]
	\begin{center}
	\includegraphics[scale=0.6]{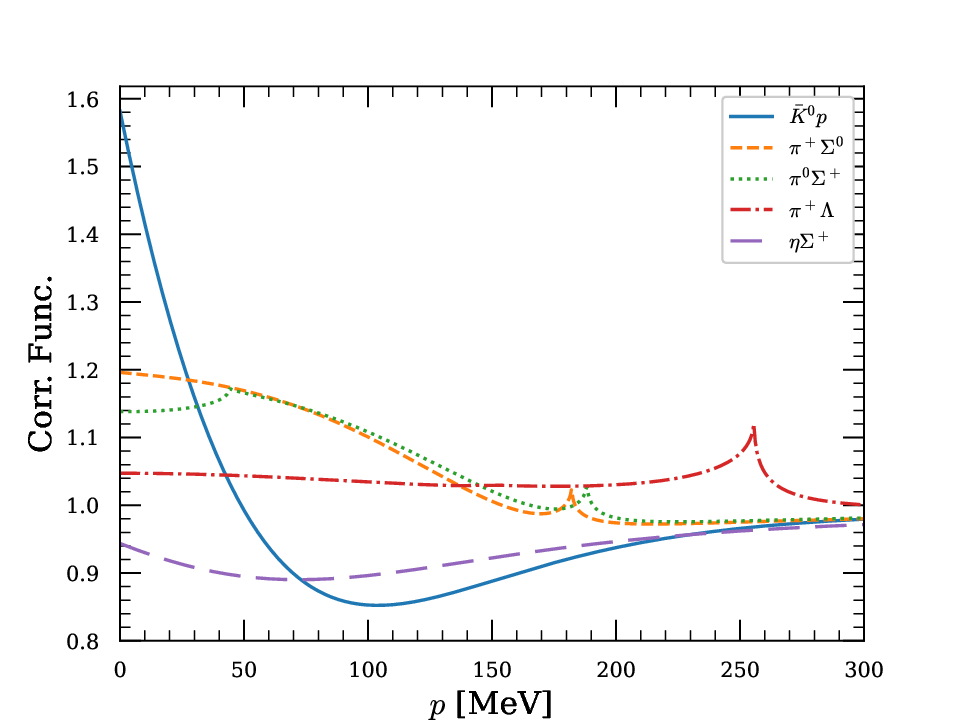}
	\end{center}
	\vspace{-0.7cm}
	\caption{Correlation functions of the different channels.}
	\label{fig:CF}
	\end{figure}

First, we use the chiral unitary approach in Sec. \ref{sec:Form} with the cutoff regularization of $q_{\text{max}}=630 \mev$.
The scattering length $a_i$ and effective range $r_i$ for different channels are presented in Table \ref{tab:ai} and Table \ref{tab:r0}, respectively.
The results for the correlation functions of the different channels are shown in Fig \ref{fig:CF}, calculated with $R=1\, \rm fm$.
The observed cusps correspond to the opening of some channels.
It is a bit surprising to see effective ranges so large in the $\pi^0 \Sigma^+, \pi^+ \Sigma^0$ and $\pi^+ \Lambda$ channels.
We have checked that in the $\pi^0 \Sigma^+, \pi^+ \Sigma^0$ channels they are related to small cusps appearing in the corresponding diagonal elements of the $T$ matrix.
Such subtle threshold effects can be much model dependent and it would be good to see if they are supported by experimental studies.

\begin{table}[t]
	\caption{Scattering length $a_i$ for channel $i$ [in units of fm].}
	\label{tab:ai}
\begin{tabular*}{0.49\textwidth}{@{\extracolsep{\fill}}ccc}
\hline
\hline
 $a_1$ &         $a_2$ &         $a_3$ \\
\hline
$0.452-i1.125 $&$ -0.150-i0.028 $&$ -0.116-i0.003 $\\
\hline
\hline
 $a_4$ &    $a_5$ & \\
 \hline
$ -0.045 $&$ 0.080-i0.151$& \\
\hline
\hline
\end{tabular*}
\end{table}
\begin{table}[t]
	\caption{Effective range $r_i$ for channel $i$. [in units of fm]}
	\label{tab:r0}
\begin{tabular*}{.49\textwidth}{@{\extracolsep{\fill}}ccc}
\hline
\hline
$r_1$ &          $r_2$ &          $r_3$\\
\hline
$0.043-i0.451 $&$ -35.243-i16.619 $&$ -67.401+i0.388 $\\
\hline
\hline
$r_4$& $r_5$&\\
\hline
$ -65.868 $&$ 0.308+i0.326$&\\
\hline
\hline
\end{tabular*}
\end{table}

\begin{figure*}[bp]
    \centering
    \begin{minipage}[b]{0.3\textwidth}
        \centering
        \includegraphics[width=\textwidth]{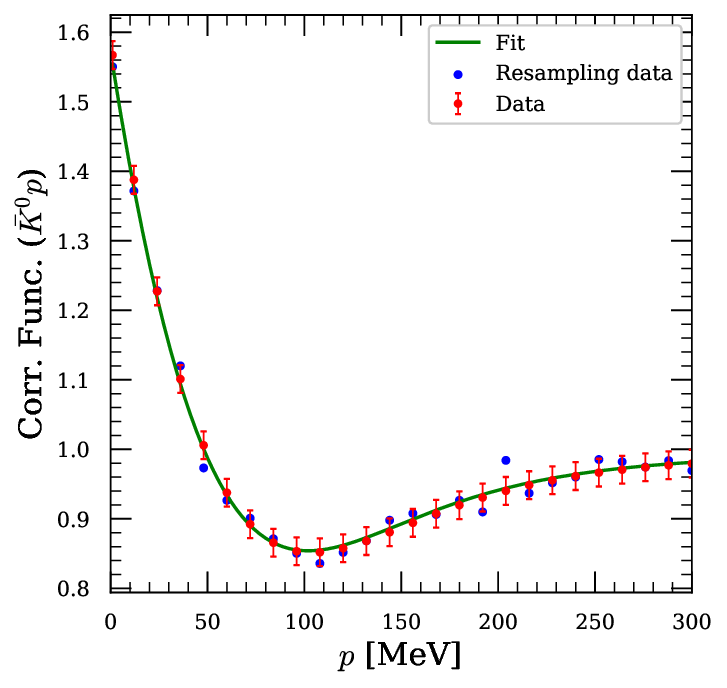}
    \end{minipage}
    \hfill
    \begin{minipage}[b]{0.3\textwidth}
        \centering
        \includegraphics[width=\textwidth]{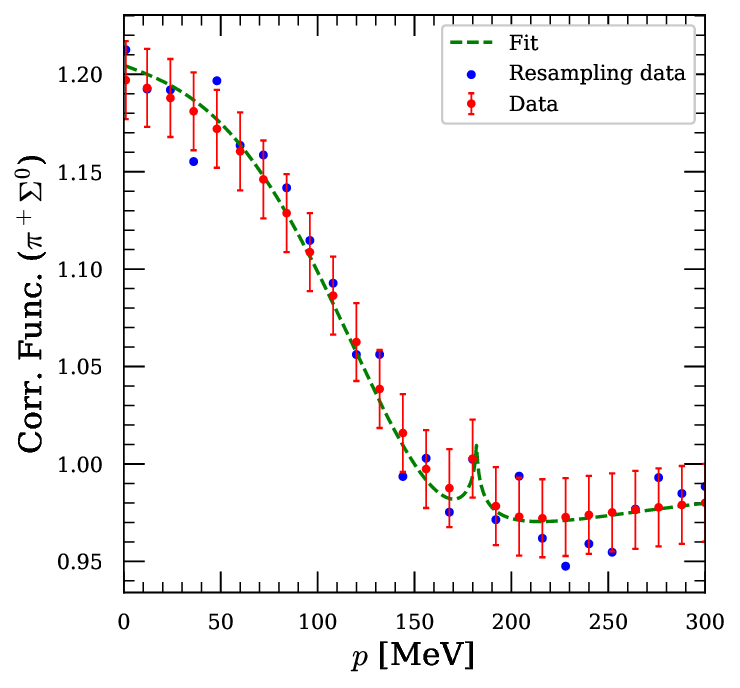}
    \end{minipage}
    \hfill
    \begin{minipage}[b]{0.3\textwidth}
        \centering
        \includegraphics[width=\textwidth]{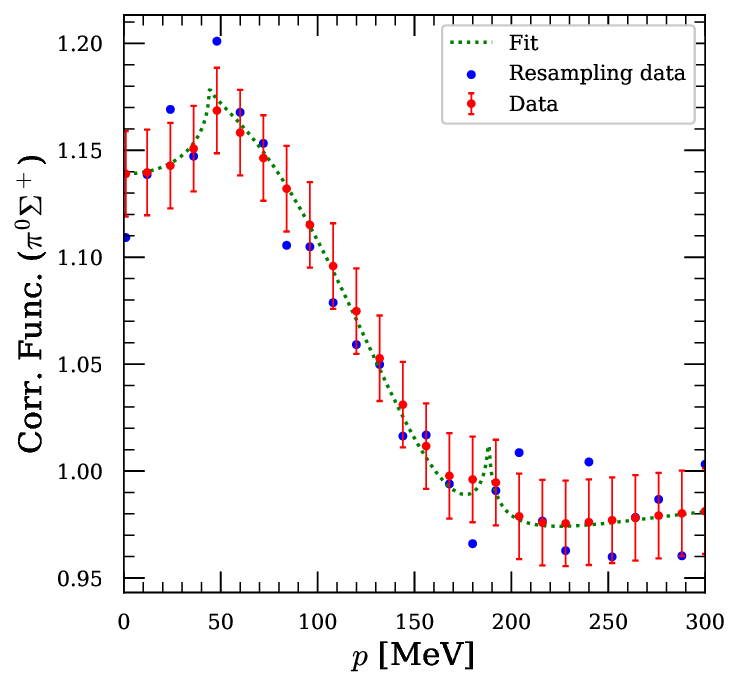}
    \end{minipage}
    \vspace{1em}
    \hspace*{\fill}
    \begin{minipage}[b]{0.3\textwidth}
        \centering
\includegraphics[width=\textwidth]{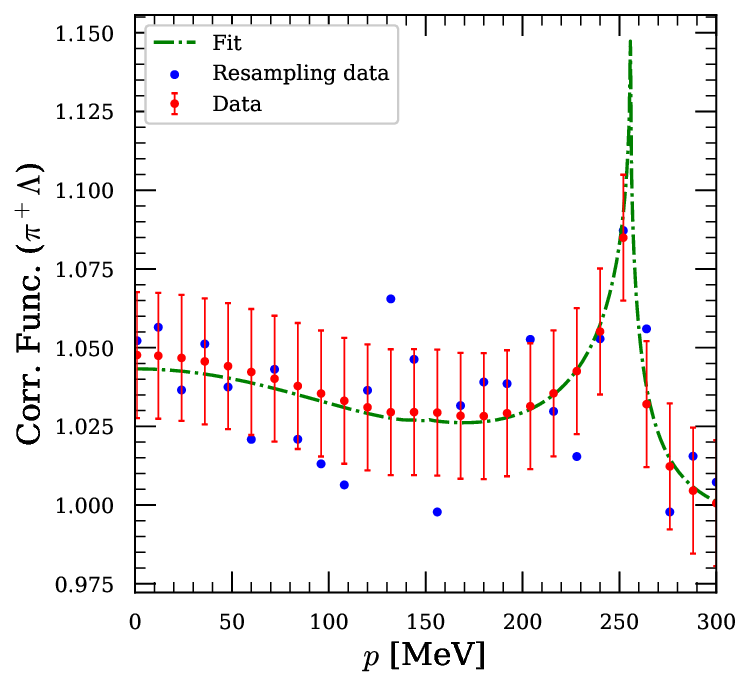}
    \end{minipage}
    \hspace{0.1\textwidth}
    \begin{minipage}[b]{0.3\textwidth}
        \centering
        \includegraphics[width=\textwidth]{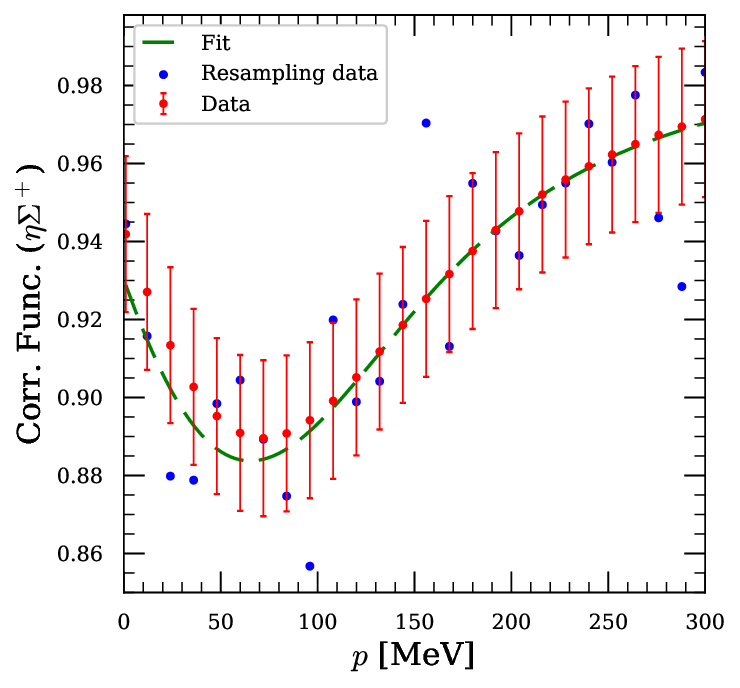}
    \end{minipage}
    \hspace*{\fill}
    \caption{Correlation functions for different channels, with $26$ points in each curve with an error of $\pm 0.02$.
	The centroids of the red data follow the theoretical curve.
	The blue point is obtained from one of the resampling runs, with a random Gaussian generation of the centroids of each point.}
    \label{fig:cf2}
\end{figure*}

Next, we discuss the results obtained from fitting the data using the resampling method.
The data with the assumed errors are shown in Fig.~\ref{fig:cf2}.
The average values and dispersion of the obtained parameters are shown in Table \ref{tab:fitpar}.
\begin{table*}[tbp]
\caption{Values obtained for the parameters.}
\label{tab:fitpar}
\begin{tabular*}{1.0\textwidth}{@{\extracolsep{\fill}}ccccc}
\hline
\hline
$\tilde{C}_{11}$ & $\tilde{C}_{12}$ & $\tilde{C}_{14}$ & $\tilde{C}_{15}$ & $\tilde{C}_{22}$ \\
	 \hline
$1.036\pm0.261$&$-0.985\pm0.138$&$-1.204\pm0.220$&$-0.829\pm0.406$&$1.924\pm0.147$\\
	\hline
	\hline
	 $\tilde{C}'_{22}$ &$\tilde{C}_{24}$&$\tilde{C}_{25}$ & $\tilde{C}_{44}$ & $\tilde{C}_{45}$\\
	\hline
	 $-2.136\pm0.465$&$-0.057\pm0.342$&$-0.028\pm0.571$&$-0.053\pm0.141$&$-0.066\pm0.706$\\
	 \hline
	 \hline
	 $\tilde{C}_{55}$ & $q_{\text{max}}$(MeV) &$R$(fm) \\
	 \hline
	 $0.043\pm0.447$&$653.468\pm63.802$&$0.995\pm0.029$\\
	 \hline
	 \hline
\end{tabular*}
\end{table*}
The first comment is that $R$ has a high precision of around $3\%$ and the $q_{\text{max}}$ follows with a precision of around $10\%$.
The matrix elements which are zero in Table~\ref{tab:cij} are small here, but the errors in $\tilde{C}_{ij}$ for subscripts $4$ and $5$ are not small.
Other elements are compatible with Table~\ref{tab:cij} within errors, which are around $20\%$ except for $\tilde{C}_{15}$.
However, one should not pay too much attention to the values of these parameters because there are correlations among them, and different sets of parameters give rise to the same results for the observables. 
This means that different sets of parameters can give the same results for the observables, and this is why the uncertainties of some parameters are so large.
This is the value of using the resampling method that randomly generates different equivalent sets of parameters and all that one has to do is to evaluate the average and dispersion of the observables obtained from the different fits in the resampling procedure.

The scattering lengths and effective ranges from the fit to the correlation functions are shown in Table \ref{tab:fita} and Table~\ref{tab:fitr}, respectively.
The scattering lengths are comparable \sout{well} to the original ones, with small errors.
The effective ranges are very large for the $\pi^0 \Sigma^+, \pi^+ \Sigma^0$ and $\pi^+ \Lambda$, and so are the uncertainties obtained from the correlation functions.
\begin{table*}[htbp]
\centering
\caption{The average and dispersion for scattering length $a_i$ for channel $i$ [in units of fm].}
\label{tab:fita}
\resizebox{1\textwidth}{!}{
\begin{tabular}{c|c|c}
\toprule\hline
$a_1$ & $a_{2}$ & $a_{3}$ \\
\hline
$(0.468\pm0.088)-i(1.130\pm0.041)$ & $-(0.148\pm0.010)-i(0.030\pm0.004)$ &$-(0.113\pm0.010)-i(0.004\pm0.003)$\\
\hline
$a_{4}$ & $a_{5}$ & \\
\hline
$-(0.045\pm0.008)$ & $(0.083\pm0.010)-i(0.161\pm0.026)$ & \\
\hline
\bottomrule
\end{tabular}
}
\end{table*}
\begin{table*}[htbp]
\centering
\caption{The average and dispersion for effective range $r_{i}$ for channel $i$ [in units of fm].}
\label{tab:fitr}
\resizebox{1\textwidth}{!}{
\begin{tabular}{c|c|c}
\toprule\hline
$r_1$ & $r_{2}$ & $r_{3}$ \\
\hline
$(0.025\pm0.150)-i(0.452\pm0.089)$ & $-(38.019\pm6.345)-i(16.534\pm1.932)$ &$-(75.053\pm17.150)+i(1.143\pm1.456)$\\
\hline
$r_{4}$ & $r_{5}$ & \\
\hline
$-(75.035\pm19.508)$ & $(0.334\pm0.761)+i(0.380\pm0.947)$ & \\
\hline
\bottomrule
\end{tabular}
}
\end{table*}

\begin{figure}[htbp]
\begin{center}
\includegraphics[scale=0.55]{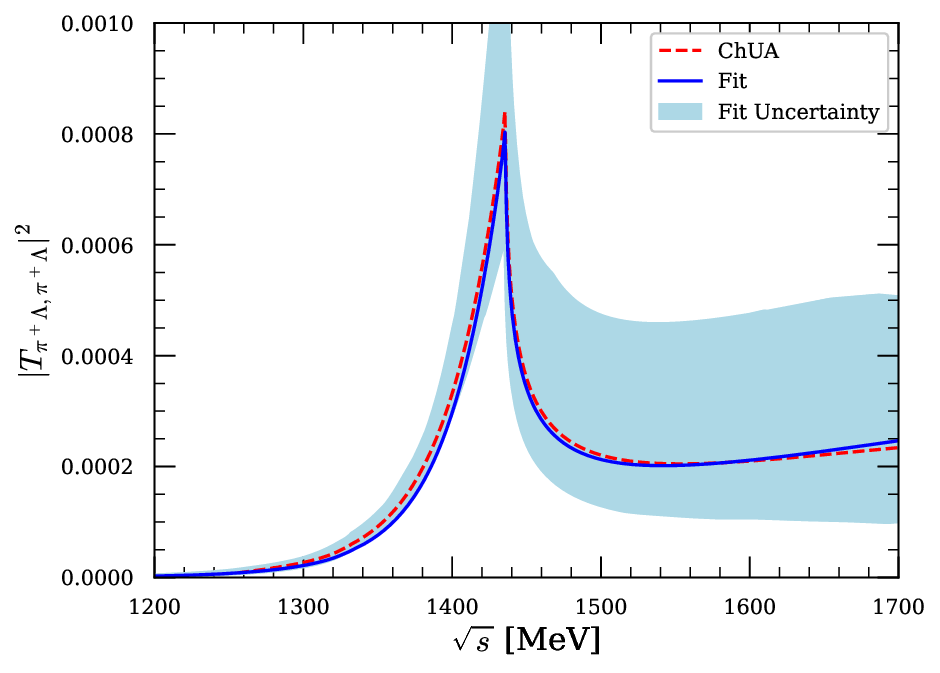}
\end{center}
\vspace{-0.7cm}
\caption{Comparison of the fit result with the one of the chiral unitary approach taking $q_{\text{max}}=630$ MeV.}
\label{fig:t2}
\end{figure}

Finally, in Fig.~\ref{fig:t2}, we show $\abs{T_{\pi^+ \Lambda \to \pi^+ \Lambda}}^2$ considering the uncertainties induced by the errors in the correlation functions.
Independently on whether there is or not a pole, the correlation functions of Fig.~\ref{fig:cf2} induce a cusplike structure for the $\pi^+ \Lambda \to \pi^+ \Lambda$ amplitude at the $\bar K N$
with the uncertainties reflected by the band of the figure.

\section{Conclusions}
We have studied the correlation functions of the coupled channels of $I=1$, $\bar K^0  p, \pi^+ \Sigma^0, \pi^0 \Sigma^+, \pi^+ \Lambda$, and $\eta \Sigma^+$, based on the chiral unitary approach, 
which produces the two $\Lambda(1405)$ states and the $\Lambda(1670)$ in $I=0$, the $\Sigma(1620)$, and surprisingly, a new state, a $\Sigma^*(1/2^-)$ state at the $\bar K N$ threshold, which, after many attempts to be found experimentally, has been clearly observed in a recent experiment by the Belle Collaboration.
We discussed that the state is in the border line of being a bound state or a virtual state, but we also show that scattering amplitudes have a smooth transition from one to the other case.

In order to learn more about this state we suggest to measure correlation functions of all the channels involved.
We have evaluated these correlation functions and then we tackled the inverse problem, 
which is to see which information we can get from the correlation functions concerning scattering parameters, and in particular, 
if from the correlation functions we can get evidence of the existence of this $\Sigma^*(1/2^-)$ state.
The answer is positive and we show that the correlation functions have encoded information leading to a clear cusp structure in $I=1$ at the $\bar K N$ threshold.

We study in detail the uncertainties in the observables that we determine from the correlation functions assuming errors typical of present measurements.
For this we use the resampling method generating many Gaussian random sets of centroids of the correlation functions data, and carrying fits to these data with a minimal model dependent approach.
The uncertainties in the scattering lengths and effective ranges of the different channels are relatively small, of the order of $20 \%$ or smaller.
And the $\Sigma^*(1/2^-)$ state at the threshold of $\bar K N$ can be induced from the minimal model  dependent framework used to fit the correlation functions. 
With improved precision in the correlation data, one could even distinguish between the state being a bound or a virtual state, although we also stress that this distinction has no repercussion on the observable magnitudes.

\section*{Acknowledgments}
E.W. acknowledges the support from the National Key R\&D Program of China (No. 2024YFE0105200).   
This work is partly supported by the National Natural Science Foundation of China (NSFC) under Grants No. 12365019, No. 11975083, No.12175239, No. 12221005, No. 12475086 and No. 12192263, 
and by the Central Government Guidance Funds for Local Scientific and Technological Development, China (No. Guike ZY22096024),
the Natural Science Foundation of Guangxi province under Grant No. 2023JJA110076,
and partly by the Natural Science Foundation of Changsha under Grant No. kq2208257 and the Natural Science Foundation of Hunan province under Grant No. 2023JJ30647 (CWX),
and the Natural Science Foundation of Henan under Grant No. 232300421140 and No. 222300420554.
This work is also partly supported by the Spanish Ministerio de Economia y Competitividad (MINECO) and European FEDER funds under Contracts No. FIS2017-84038-C2-1-P B, PID2020-112777GB-I00, and by Generalitat Valenciana under contract PROMETEO/2020/023.
This project has received funding from the European Union Horizon 2020 research and innovation programme under the program H2020-INFRAIA-2018-1, Grant Agreement No. 824093 of the STRONG-2020 project, and by the Xiaomi Foundation / Xiaomi Young Talents Program.

\bibliographystyle{a}
\bibliography{refs}
\end{document}